\documentclass{wiley-article}

\usepackage{siunitx}

\usepackage[linesnumbered,boxed,ruled,commentsnumbered]{algorithm2e}
\usepackage{algpseudocode}
\usepackage{lineno}
\usepackage{amsmath,amssymb,amsfonts}
\usepackage{graphicx}
\usepackage{textcomp}
\usepackage{xcolor}
\usepackage{listings}
\usepackage{booktabs}
\usepackage{array}
\usepackage{ifpdf}
\usepackage{mdwmath}
\usepackage{mdwtab}
\usepackage{multirow}
\usepackage{xspace}
\usepackage{url}
\usepackage{color,soul}
\usepackage[utf8]{inputenc}
\usepackage[export]{adjustbox}
\def\toolname{Lorikeet\xspace}

\PassOptionsToPackage{hyphens}{url}\usepackage[pdftex, colorlinks=true, hyperfootnotes=false, hyperindex=true, plainpages=false, pagebackref=false, pdfpagelabels=true, pdfstartview=FitH, linkcolor=blue, citecolor=blue, urlcolor=blue]{hyperref}

\usepackage[textsize=footnotesize,backgroundcolor=yellow!40]{todonotes}

\papertype{Original Article}
\paperfield{Journal Section}

\title{Integrated Model-Driven Engineering of Blockchain Applications for Business Processes and Asset Management}

\author[1]{Qinghua Lu}
\author[3\authfn{1}]{An Binh Tran}
\author[2\authfn{1}]{Ingo Weber}
\author[1]{Hugo O'Connor}
\author[3\authfn{1}]{Paul Rimba}
\author[1]{Xiwei Xu}
\author[1]{Mark Staples}
\author[1]{Liming Zhu}
\author[1]{Ross Jeffery}

\contrib[\authfn{1}]{Majority of the work done while this author was with Data61, CSIRO.}

\affil[1]{Data61, CSIRO, Sydney, Australia\\ firstname.lastname@data61.csiro.au}
\affil[2]{Technische Universitaet Berlin, Germany\\ingo.weber@tu-berlin.de}

\affil[3]{Deputy, Sydney, Australia\\atran@deputy, primba@deputy.com}

\corraddress{Qinghua Lu\\Data61, CSIRO, Sydney, Australia}
\corremail{qinghua.lu@data61.csiro.au}

\runningauthor{Qinghua Lu et al.}

\begin{document}

\maketitle

\begin{abstract}
Blockchain has attracted broad interests to build decentralised applications. 
A typical class of applications uses blockchain for the management of cross-organisational business processes as well as assets. 
However, developing such applications without introducing vulnerabilities is hard for developers, not the least because the deployed code is immutable and can be called by anyone with access to the network. 
Model-driven engineering (MDE) helps to reduce those risks, by combining proven code snippets as per the model specification, which is easier to understand than source code. 
Therefore, in this paper, we present an approach for integrated MDE across business processes and asset management (e.g. for settlement). 
Our approach includes methods for fungible/non-fungible asset registration, escrow for conditional payment, and asset swap. 
The proposed MDE approach is implemented in a smart contract generation tool called Lorikeet, and evaluated in terms of feasibility, functional correctness, and cost effectiveness.

\end{abstract}

\section{Introduction}
Blockchain has attracted a wide range of interests from start-ups, enterprises and governments. 
Those interests have been sparked by the possibility of using blockchain as a general, decentralised and trustworthy computing environment through the advent of smart contracts. 
A large number of projects have been conducted to explore how to use blockchain to re-architect systems and to build new applications and business models~\cite{2019-Bratanova-ACS}.

A typical class of applications uses blockchain for the management of business processes across organisations as well as for digital asset management, which are maintained and controlled on-chain. 
Assets can be classified into fungible assets and non-fungible assets. Fungible assets are individual units that are interchangeable (e.g., company share and gold), while non-fungible assets represent unique assets (e.g., cars, patents, houses). Both fungible and non-fungible assets are traditionally managed by relying on a centralised trusted authority, which can cause trust issues and introduce inefficiencies or counterparty risks (e.g., re-assigning ownership of goods before payment).
\begin{figure}[t]
	\begin{center}
		\includegraphics[width = 0.60\columnwidth]{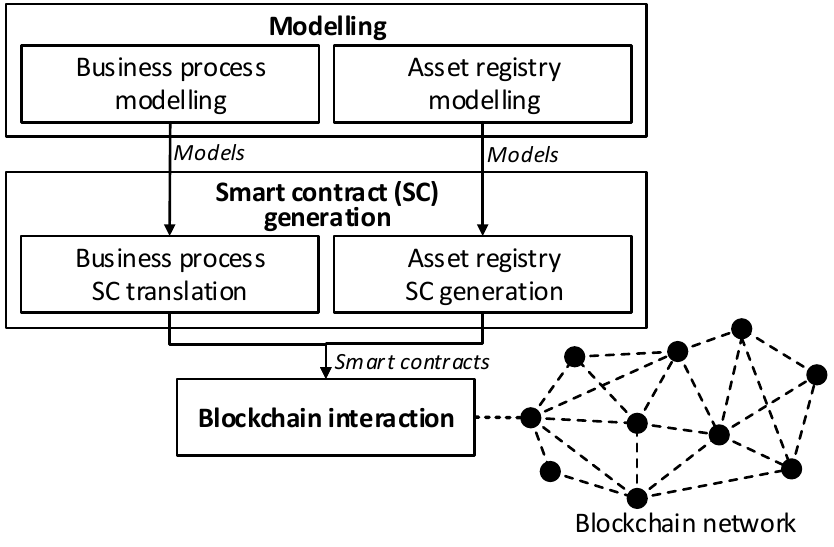}
		\caption{Overview of our MDE approach.}
		\label{overview}
	\end{center}
\end{figure}
However, it is hard for developers to develop blockchain applications for business processes and asset management without introducing vulnerabilities or bugs, not the least because the deployed smart contract code is immutable and can be called by anyone with access to the network \cite{Luu:2016:MSC:2976749.2978309}. 
Model-driven engineering (MDE) \cite{MODELS2017} helps to reduce those risks, by combining proven code snippets as per the model specification, which is typically easier to understand than source code with all its implications.

Our previous efforts targeted MDE for business processes \cite{Weber:BPM2016} and non-fungible asset registries \cite{Tran2017} in isolation. Fungible asset registries were not addressed in~\cite{Weber:BPM2016,Tran2017}. Also, integration of business processes with registries is required for asset management on blockchain since registering assets is often not as simple as only storing asset records. For example, the authoritative organisation(s) may need to collect documents and check if the application is valid. Conversely, business processes that touch on asset registries, e.g., for creating assets or changing their ownership, need to be integrated with registries. Therefore, in this paper, as shown in \autoref{overview}, we present an approach for integrated MDE across business processes and asset management: business processes are supported, as are fungible (e.g., ERC20 tokens) and non-fungible (e.g., car/grain/land titles) assets. Business processes and asset management are integrated in that business processes can control assets, and assets can make use of business processes.
The value of this integration is managing processes for both fungible and non-fungible asset registration, escrow for conditional payment, and asset swap in an efficient way using blockchain, which are not covered by our previous work.
We design and develop a tool called Lorikeet\footnote{Rainbow Lorikeet is a species of parrots often encountered in Sydney, Australia -- see \url{https://en.wikipedia.org/wiki/Rainbow_lorikeet} (accessed 26 July 2020)} that implements the proposed MDE approach and transforms models into smart contracts in Solidity which can be compiled for Ethereum~\cite{wood2014ethereum} and other blockchain platforms. We use Ethereum as the blockchain technology platform, which can well be used for any types of deployment (i.e., public/private/consortium deployments). In other words, our approach is by no means restricted to the public Ethereum network. Our evaluation results show that the proposed approach is feasible and functionally correct.



The contributions of this paper are as follows.
\begin{itemize}
	\item An MDE approach for development of blockchain applications for business processes and asset management.
	\begin{itemize}
		\item Modelling methods to specify models for integration of business processes with asset management, including both fungible and non-fungible asset registration, escrow for conditional payment, and asset swap. We provide the templates for the developers to customize data schemas for both fungible and non-fungible assets registries. We also extend the OMG standard Business Process Model and Notation (BPMN) 2.0 to specify interactions between business processes and fungible/non-fungible asset registries.
		\item Smart contract generation methods to automatically transform models into smart contract programming language - Solidity, which can be compiled for Ethereum and other blockchain platforms. The input models include business process models, and fungible/non-fungible registry data schemas, while the generated smart contracts consist of business process execution smart contracts and standardised ERC-20/ERC-721 compliant asset registry smart contracts. Interactions between business processes and asset registries (e.g., for escrow or asset swap) are also implemented in the produced smart contracts.
		\item Blockchain interaction methods to handle compilation and deployment of smart contracts and communication with the deployed smart contracts on blockchain.
	\end{itemize}
	\item Feasibility and functional correctness evaluation using four industrial use cases which cover fungible and non-fungible asset registration, escrow for conditional payment, and asset swap.
	\item An analysis of gas consumption and comparison with numbers from over 292 million transactions on the public Ethereum blockchain.
\end{itemize}

The remainder of this paper is organised as follows. \autoref{sec:bg} discusses the background and related work. \autoref{sec:approach} presents our MDE approach. \autoref{sec:tool} introduces our tool named Lorikeet. \autoref{sec:eval} evaluates the proposed approach using use cases. \autoref{sec:concl} concludes the paper and outlines the future work.

\section{Background and Related Work}
\label{sec:bg}
In this section, we first introduce blockchain technology and smart contracts in \autoref{sec:bg_bc}. Then, we provide background knowledge of Model-Driven Engineering (MDE) and its benefits in \autoref{sec:bg_mdd}.  Finally, we explain why MDE and blockchain can be a solution for addressing the trust issue in the business process domain (\autoref{sec:bg_bp}) and the asset registry domain (\autoref{sec:bg_reg}), which has not been fully solved before.

\subsection{Blockchain and Smart Contracts}
\label{sec:bg_bc}
A \emph{blockchain} is an append-only store of transactions distributed across computational nodes and structured as a linked list of blocks, each containing a set of transactions. The main purpose of structuring the data store into blocks is to obtain manageable chunks of information, for communication as well as for achieving consensus.
Blockchain was introduced as the technology behind Bitcoin \cite{Satoshi:bitcoin}. 
Its concepts have been generalized to \emph{distributed ledger} systems that verify and store any transactions without coins or tokens \cite{scheuermann2015iacr}, without relying on any central trusted authority like traditional banking or payment systems. 
Instead, all participants in the network can reach agreements on the states of transactional data to achieve trust.

A smart contract is a user-defined program that is deployed and executed on a blockchain system \cite{Omohundro:2014}, which can express triggers, conditions and business logic \cite{Weber:BPM2016} to enable complex programmable transactions. 
Smart contracts can be deployed and invoked through transactions, and are executed across the blockchain network by all connected nodes. 
The signature of the transaction sender authorizes the data payload of a transaction to create or execute a smart contract. 
Trust in the correct execution of smart contracts extends directly from regular transactions, since (i) they are deployed as data in a transaction and thus immutable; (ii) all their inputs are through transactions and the current state; (iii) their code is deterministic; and (iv) the results of transactions are captured in the state and receipt trees, which are part of the consensus. 

When using a blockchain, there are different types of deployments, including public blockchain, consortium blockchain or private blockchain. 
Public blockchains, which can be accessed by anyone on the Internet (\emph{``permission-less''}), have high information transparency and auditability, but sacrifice performance and a cost/incentive model. 
A consortium blockchain is typically used across multiple organisations and the rights to read/write on the blockchain may be restricted to specific participants.
In a private blockchain network, write permissions are often kept within one organisation, although this may include multiple divisions of a single organisation.
Private blockchains are the most flexible for configuration because the network is governed and hosted by a single organisation.
A blockchain may be \emph{permissioned} in requiring that one or more authorities act as a gate for participation. 
This may include permission to join the network and read information from the blockchain, to initiate transactions, or to create blocks. 
Permissions can be stored either on-chain or off-chain.
There are often tradeoffs between permissioned and permission-less blockchains including transaction processing rate, cost, censorship-resistance, reversibility, finality and flexibility in changing and optimising the network rules.

Data privacy and scalability are two major challenges of public blockchains. Data privacy is limited because there is no privileged user in public blockchain: everyone can join the public blockchain, access all the data, and validate new transactions. The scalability limits on public blockchain include the size of the data, the transaction processing rate, and the latency of data transmission and commits (e.g., around 1 hour on Bitcoin and 3 minutes on Ethereum). The number of transactions included in each block is limited. The block size limit in Bitcoin is 1MB, while gas limit (gas is the pricing unit for transaction execution and data storage in Ethereum) is applied in Ethereum to limit the number and complexity of transactions to be included into a block.

\subsection{Asset management}
A typical class of blockchain applications is digital asset management~\cite{Luetal2018}. Assets can be classified into fungible assets and non-fungible assets. Fungible assets are individual units that are interchangeable. For example, \$10 notes are interchangeable with other \$10 notes and can be swapped for two \$5 notes.
Non-fungible assets represent unique assets. For example, all diamonds have different sizes, shapes, and grades. It is difficult to find two diamonds with the same value.
Similarly, patents or houses are unique, due to their intrinsic properties like location.

The core of asset management comprises the processes of registering and transferring asset ownership in accordance with the terms of an underlying contract. This includes asset registry, escrow for conditional payment and asset swap, as follows. Asset registry maintains a list of assets that belong to a party. Escrow is a financial arrangement where a third party holds and regulates payment of the funds or other assets on behalf of two parties involved in a transaction. Payment is kept in an escrow account and only released when all the obligations of an agreement are fulfilled. Asset swap is the ideally atomic exchange of assets based on an amount agreed by both parties of the transaction. 

Assets are traditionally managed by relying on a centralised trusted authority, which can cause trust issues and introduce inefficiencies or counterparty risks (e.g., re-assigning ownership of goods before payment, expensive escrow fees, etc.). Blockchain can replace the centralised trusted authority and provide a decentralised trusted infrastructure to maintain asset registries and facilitate asset transactions. To improve asset liquidity, Ethereum Request for Comment (ERC) \footnote{\url{https://docs.ethhub.io/built-on-ethereum/erc-token-standards/what-are-erc-tokens}} defines a set of standard development interfaces that Ethereum-based tokens' (i.e., assets) smart contracts must comply with. ERC-20\footnote{\url{https://eips.ethereum.org/EIPS/eip-20}} and ERC-721\footnote{\url{https://eips.ethereum.org/EIPS/eip-721}} are the most popular Ethereum token standards for fungible and non-fungible assets respectively.

\subsection{Model-Driven Engineering}
\label{sec:bg_mdd}
Model-driven engineering (MDE) is a methodology that uses models at various levels of abstraction to address software development complexity \cite{Schmidt2006}. Domain-specific MDE can help map the model of the problem domain to the design of the software solution \cite{Evans:2003:DDT:861502,Fowler:2010:DSL:1809745}. The abstraction level for models can be at various degrees. For example, some models can directly derive the production code while others can only be used to guide the developers in developing the software. In model-driven engineering, models can produce code or guide implementation, or conversely, the code (or other artefacts) can generate models to help understand the software design, e.g.\ database schema. Depending on the purpose, various concepts can be captured in models, e.g.\ system/database structures or a sequence of activities. 

Specifically, MDE for code generation can be further classified into different types: once-off code generation and repetitive code generation. Once-off code generation means that once the code is derived from the model, the subsequent evolution of the code is independent of the model, whereas in repetitive code generation, the code is re-generated from the model following subsequent changes to the model over time. The repetitive code generation can be further classified into one-way model-to-code code generation and round-trip code generation. In one-way model-to-code code generation, the code is updated if changes are made to the model, but not vice versa. While in round-trip code generation, if the generated code is updated, the changes can be propagated back to the model level. 

In the context of blockchain-based applications, MDE is of particular importance for the following reasons \cite{Luetal2018}. First, model-driven engineering tools can implement best practices and generate well-tested code, thereby avoiding vulnerable code which may be easily attacked (e.g.\ the DAO exploit on the Ethereum blockchain\footnote{\url{http://www.coindesk.com/understanding-dao-hack-journalists}}). Second, models can avoid lock-in to specific blockchain technologies since they can be platform-agnostic, and a model-driven engineering tool might be able to produce artefacts for multiple blockchain platforms. Third, models are easier to understand than code, thus improving the development productivity. It is easier to check the correctness of a model and MDE tools can ensure that the deployed code has not been modified after its generation from the model. Fourth, it can facilitate communication with domain experts since domain experts can look at the model to understand how their ideas are represented in the system.

\subsection{Business Processes}
\label{sec:bg_bp}
Trust issues in business processes \cite{viriyasitavat2011relation} have been discussed over the last decade. \cite{carminati2014secure} uses selective encryption and restricts data access for both the broker and the service partners to achieve trust with untrusted broker. Mont and Tomasi \cite{mont2001distributed} design a trust service for cross-company collaboration based on a hybrid architecture mixing a trusted centralised control with untrusted peer-to-peer components. \cite{li2010distributed} present an agent-based architecture that can remove the scalability bottleneck of a centralised orchestration engine and provides more efficiency by executing portions of processes close to the data they operate on. \cite{SquicciariniPB08a} select partners on the basis of disclosure policies and credentials (i.e. identity attributes issued by a ``Credential Authority'') in virtual organisations. Various important concepts such as conformance \cite{Aalst2008}, reliability \cite{Subramanian2008} and quality of services \cite{Zeng2004} have been studied for centrally controlled business process execution. However, these works do not solve the trust issue as a collaborating party might have corrupted their historic files to their advantage. There is no party that sees all the messages in the business processes.

To address the trust issue, similar to our previous work~\cite{Weber:BPM2016}, blockchain has been adopted to define high-level business process models in smart contracts that are deployed and executed on blockchain~\cite{Luciano2017, Nakamura2018} without taking into account process instance data.
The tool Caterpillar \cite{Lopez-Pintado:BPM2017, DiCiccio2019} can support both control flow and instance data using blockchain. The extension of Caterpillar adds the feature of supporting runtime adaptation of a business process~\cite{8944990}. However, Caterpillar is a Business Process Management System (BPMS) operating on blockchain, and does not support asset management.

In this paper, our work focuses on using blockchain to address the trust issue in both the business process domain and the asset registry domain. Lorikeet is an MDE tool that addresses the challenge of integrating asset registry with business processes on blockchain.

\subsection{Registries}
\label{sec:bg_reg}
A registry is a list of data recorded and maintained by a trusted authority, which is an authoritative database for specific entities and is used to manage many aspects of daily life, such as land titles, business names, books, marriages, births and deaths, music, films and domain names. Traditionally, registries are maintained by a central authority. However, such centralised architecture may cause a single point of failure for the whole registry system. Building registries on a blockchain can guarantee data integrity, availability, transparency and immutability, which are key requirements for registries \cite{Downey2016}. Additionally, blockchain can be used as a unified infrastructure which enables multiple registries to easily interact with each other. 

There are registries being built on blockchain in ad-hoc ways, for example, Namecoin\footnote{\url{https://namecoin.org/}}, which is a domain name registry that shares the same network with Bitcoin\footnote{\url{https://bitcoin.org/}}, and Abscribe\footnote{\url{https://www.ascribe.io/}}, which is an artwork registry that enables artists to maintain the ownership of their digital artwork. However, building a registry on blockchain is challenging since developers need to understand in depth how particular blockchain platforms operate and learn smart contract programming languages. Regis\footnote{\url{https://regis.nu/}} is a smart contract generation tool on Ethereum\footnote{\url{https://www.ethereum.org/}} blockchain, but only provides basic operations. We introduced our registry generator tool for blockchain in \cite{Tran2017} and briefly discussed how to integrate registries with business processes in a demo paper~\cite{Tran2018LorikeetAM}.

\section{An Integrated Model-Driven Blockchain Application Development Approach for Business Processes and Asset Management}
\label{sec:approach}
In this section, we present our Model-driven blockchain application development approach for business processes and asset management. We first provide an overview of our MDE approach in Section 3.1. Then, we discuss the modelling methods proposed for fungible/non-fungible asset registries and extensions of BPMN to support modelling of interactions between business processes and asset registries in Section 3.2. After that, we propose the methods for business process smart contract translation and registry smart contract generation in Section 3.3. Finally, we explain the blockchain interaction methods for connecting with a blockchain node, and handling the compilation, deployment as well as communication with smart contracts in Section 3.4.

\subsection{Overview of the Model-Driven Engineering Approach}
\begin{figure}[t]
	\begin{center}
		\includegraphics[width = 0.95\columnwidth]{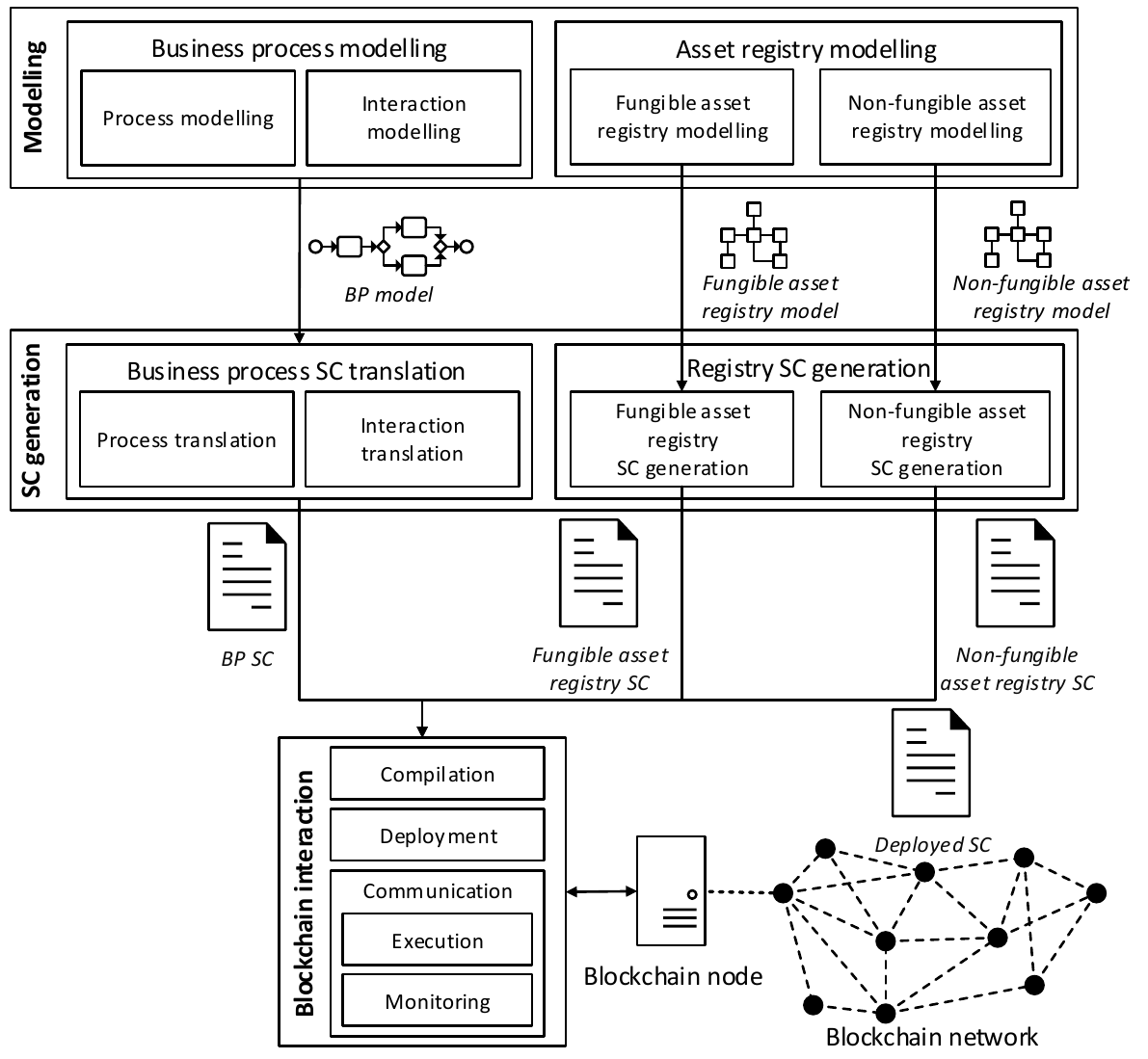}
		\caption{Architecture of our MDE approach.}
		\label{approach}
	\end{center}
\end{figure}

\autoref{approach} illustrates an overview of our model-driven engineering (MDE) approach for integrating business processes with asset management on blockchain. The design of the approach consists of three parts: modelling, smart contract (SC) generation, and blockchain interaction. For modelling, the approach provides templates for the developers to customize fungible/non-fungible asset registry data schemas and extends BPMN 2.0 to support modelling of interactions between business processes and asset registries (e.g.\ for fungible/non-fungible asset registration, escrow for conditional payment, and asset swap). The approach then transforms the built models (i.e. models for business processes, asset registries, and their integration) into blockchain smart contract implementations in a programming language (such as Solidity) and handles interaction with smart contracts deployed on blockchain.
There are three types of users for this MDE approach (and the corresponding tool): 1) developers can use it to improve development productivity and quality, 2) operators can use it to monitor the execution of generated smart contracts; 3) and domain experts can use it to communicate with developers and understand how their ideas are represented in the system.

\begin{table}[t]
	\footnotesize
	\centering
	\caption{Custom BPMN Elements}
	\label{tab:BPMN}
	\smallskip
	\begin{tabular}{@{}p{5.3cm}p{4.3cm}p{1.6cm}@{}}
		\toprule
			Element 		  & Description & Notation \\ 
			\midrule
			\vspace*{-8pt}\textit{bcext} & \vspace{-8pt}BPMN meta-model name space & \vspace{-8pt}N/A \\ \midrule
			\vspace*{-4pt}\textit{SmartContractInterface} & \vspace{-4pt}Smart contract element & \vspace{-12pt}\includegraphics[valign=T, width=0.08\textwidth]{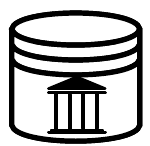} \\ \midrule
			\vspace{-5pt}\textit{ConnectionOutgoingContractInvocation} & \vspace{-5pt}Connection to the smart contract  & \vspace{-10pt}\includegraphics[valign=T]{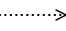} \\ 
		\bottomrule
	\end{tabular}
\end{table}

\subsection{Modelling}
As shown in \autoref{approach}, we propose modelling methods for asset registries, business processes, and their interactions. For registry modelling, we provide different methods for modelling fungible assets (e.g.\ ERC20 tokens) and non-fungible assets (e.g.\ car/grain/land titles) respectively. Business process modelling includes process modelling in BPMN 2.0 and interaction modelling using the newly extended BPMN elements. 


\subsubsection{Registry Modelling}
\begin{figure}[t]
	\begin{center}
		\includegraphics[width = 0.85\textwidth]{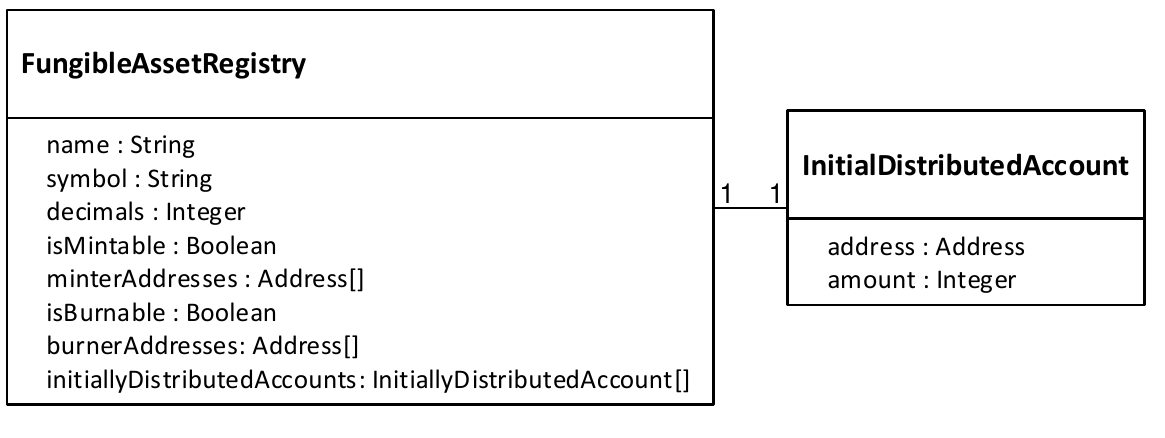}
		\caption{Fungible asset registry data model.}
		\label{fungible}
	\end{center}
\end{figure}

On the registry side, we provide modelling methods for users to design fungible and non-fungible assets via the respective data registry template in UML. The fungible and non-fungible assets are represented using different data registry templates since non-fungible assets require customised definition of attributes to describe uniqueness. Also, the templates ensure that the resulting registries comply with ERC-20/ERC-72 standards to facilitate development and, due to wide-spread use of these de-facto standards, offer high liquidity. \autoref{fungible} shows the data model for fungible asset registry, which consists of basic token details and advanced token features. The basic token details include token \textit{name}, \textit{symbol} (an abbreviation, like ``ETH'' for Ether, usually 3 or 4 characters in length), and \textit{decimals} (the number of digits in the fraction part). 

The advanced features describe token design details about minting (\textit{isMintable}, \textit{minterAddresses}), burning (\textit{isBurnable}, \textit{burnerAddresses}), and initial distribution (\textit{initiallyDistributedAccounts}). Users can configure the accounts that can mint or burn the token, the total supply of the token (the total number of tokens that have been or will be mined), and the accounts that receive the initial distribution and the amount sent to each account. 

A user can design a registry model for a new fungible token by providing the above inputs. For example, the token \textit{name} can be \textit{Lorikeet Coin}, while the \textit{symbol} is \textit{LRK}. The \textit{decimals} value can be set to \textit{2}. If isMintable and isBurnable are set to \textit{true}, the user needs to add the respective account addresses of accounts which are allowed to create or destroy tokens, respectively. Finally, the user can set up the \textit{initiallyDistributedAccounts} to identify which account addresses will receive which share of the initial distribution of \textit{Lorikeet Coin}. 


\begin{figure}[t]
	\begin{center}
		\includegraphics[width = 0.75\textwidth]{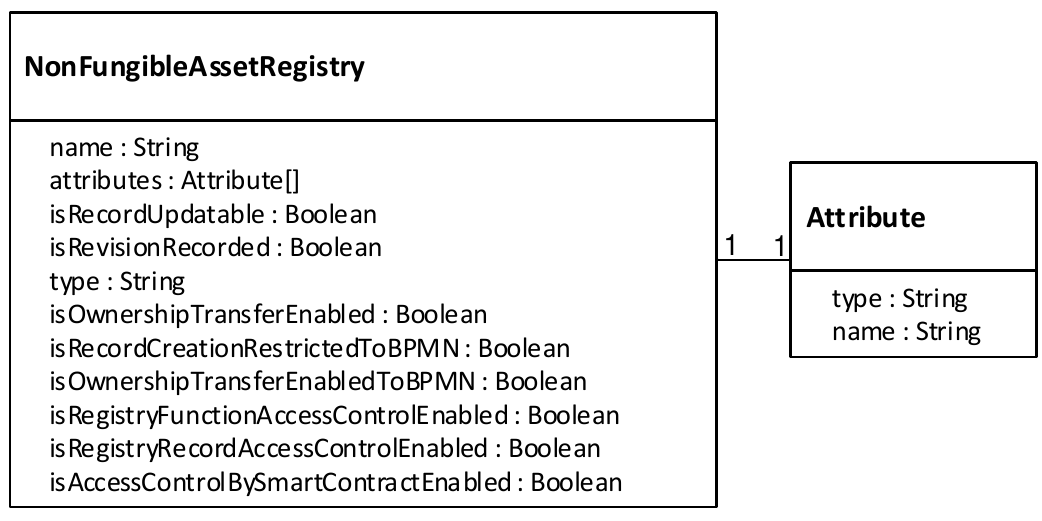}
		\caption{Non-fungible asset registry data model.}
		\label{nonfungible}
	\end{center}
\end{figure}

\autoref{nonfungible} illustrates the data structure for non-fungible asset registry, which specifies basic information and advanced features of non-fungible assets. Basic information includes registry \textit{name}, \textit{type}, and user-defined \textit{attributes}. Registry \textit{type} can be `single' or `distributed'. The `single' registry type holds all records as values in the data store as a singleton registry smart contract, which is suitable for simple registries. The `distributed' registry type manages each record as a separate smart contract, which is suitable for registries with complex operations, such as individual record-level permission management. A main registry smart contract creates these contracts and stores pointers to them. Regarding user-defined attributes, users can specify attribute name, type, whether a record is updatable, and maintain a detailed history of changes made to every registry record.

The non-fungible asset registry modeller also supports advanced features including record lifecycle management and access control. 
The registry record lifecycle (create, read, update, and delete) can be enabled to be managed via a business process executed on the blockchain. 
In this case, only the business process instance is allowed to create or update records, even though the registry is readable by the public (\textit{isOwnershipTransferEnabled}, \textit{isRecordCreationRestrictedToBPMN}, \textit{isOwnershipTransferEnabledToBPMN}). 
For example, transferring ownership can be restricted to a business process instance, which exchanges ownership of the registered items, such as grain title. 
Access control can be enabled on registry functions (\textit{isRegistryFunctionAccessControlEnabled}) and individual registry records (\textit{isRegistryRecordAccessControlEnabled}), which can be implemented within the registry smart contract or using a separate smart contract (\textit{isAccessControlBySmartContractEnabled}).

A user can model a non-fungible asset by using the features listed in Fig.\ref{nonfungible}. For example, to model a registry of grain title, on the model side, the user can fill in \textit{GrainTitle} as the \textit{name} and select \textit{single} as the \textit{registry type}. For attributes of \textit{GrainTitle}, \textit{weight} and \textit{quality} can be set as the \textit{attribute name} and units be selected as the \textit{type}. Grain title ownership can be transferred by setting \textit{isOwnershipTranferEnabledToBPMN} as \textit{true}. The user can enable \textit{isRecordCreationRestrictedToBPMN} as \textit{true} to specify that grain title creation and transfer will be managed by external BPMN processes.

\subsubsection{BPMN Modelling}
In BPMN modelling, there are three types of tasks we use here, namely default tasks, user tasks, and script tasks. All three types of tasks are implemented as functions in the process smart contract generated. Script tasks are executed automatically when they are activated. All other tasks can be executed by calling the respective smart contract function (e.g., through UI elements generated). As such, default tasks behave from the smart contract point of view like user tasks. The differentiation is only on the model level, e.g., to allow differentiating tasks that are invoked by people as opposed to off-chain software components. For user tasks, user task input parameters can be bound to registry action's input. For example, grain title ID and farmer account are provided as input in User Task ``Create Grain Title'' and bound to \textit{record\_id} and \textit{owner}. The user can write scripts to be executed in script tasks through the task templates. For example, for script task ``Calculate Grain Weight'', the consignment weight is calculated based on the truck weight with consignment and truck weight without consignment, which is represented as \textit{consignmentWeight=truckWeightWithConsignment-truckWeightWithoutConsignment}. 

In addition to standard BPMN modelling, model-level integration of business processes with asset management requires interface specification of smart contracts including asset/data registry 
smart contracts and escrow smart contracts. A smart contract on a blockchain, among others, acts as a data store which tasks in a business 
process can read data from, or write data to. However, the existing BPMN 2.0 
elements (i.e. \textit{DataStoreReference} and \textit{DataOutputAssociation})
 do not support representation of properties specific to smart contracts (e.g., registry smart contract addresses and smart contract invocation). Thus, 
to support integration, we introduce custom elements to BPMN 2.0 and design 
respective graphical representations for them, which are shown in \autoref{tab:BPMN}. The 
elements include \textit{bcext}, \textit{SmartContractInterface}, \textit{
ConnectionOutgoingContractInvocation}. \textit{bcext} is the custom namespace for blockchain smart contract relevant BPMN meta-model. \textit{SmartContractInterface} represents an interface for a smart contract which is 
external to the business process. \textit{ConnectionOutgoingContractInvocation} 
is the custom connection which links business process tasks with the 
external smart contract in the BPMN model. The graphical notation for \textit{SmartContractInterface} is extended from the existing BPMN notation for \textit
{DataStoreReference}, while the graphical notation for \textit{ConnectionOutgoingContractInvocation} is designed extending the current BPMN 
notation \textit{DataOutputAssociation}. 

\begin{figure}[t]
	\begin{center}
		\includegraphics[width = \columnwidth]{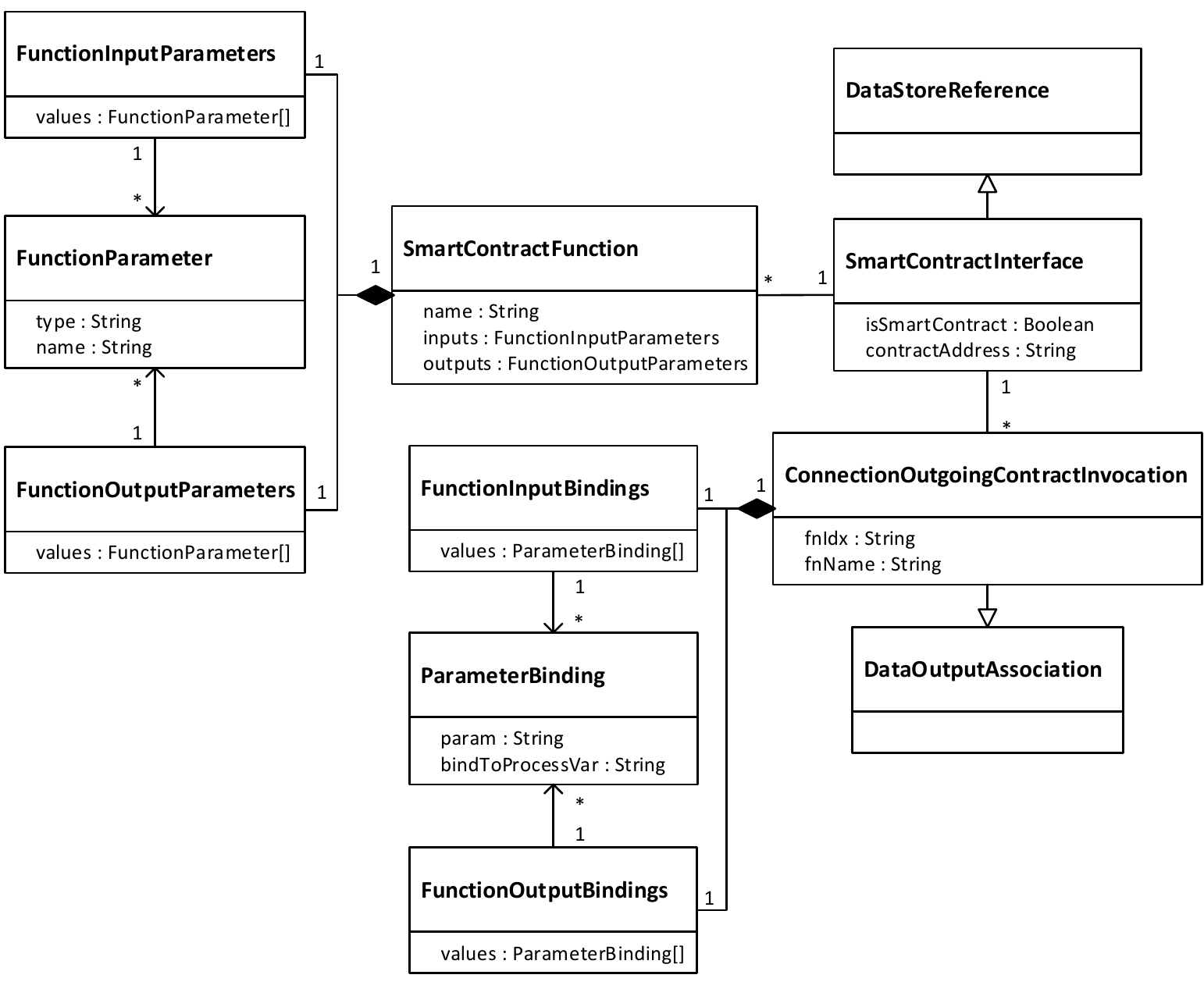}
		\caption{Meta-Model for Smart Contract Interfaces.}
		\label{classdiagram}
	\end{center}
\end{figure}

\autoref{classdiagram} shows the data structure of the proposed BPMN meta-model for smart contract-relevant extensions. 
\textit{SmartContractInterface} can represent any type of smart contracts, e.g., asset/data registry smart contract, escrow smart contract, etc.
When a smart contract is deployed on blockchain, it is uniquely identified and reachable via a smart contract address. 
In order for the business process smart contract to interact with the smart contract, the smart contract address can be made available to the BPMN model via the attribute \textit{contractAddress} in \textit{SmartContractInterface}.
If the smart contract address is provided, the address is fixed for all the instances of this business process and cannot be changed. 
If it is not provided in the model, the translated business process smart contract allows users to specify the registry contract address each time a new business process instance is created. 
Each \textit{SmartContractInterface} can have multiple \textit{SmartContractFunction}s, which are provided to inform the BPMN model of which functions are available to interact with a given smart contract and how to invoke each smart contract function. 
Each \textit{SmartContractFunction} has one of each \textit{FunctionInputParameters} and \textit{FunctionOutputParameters}. Each of \textit{FunctionInputParameters} and \textit{FunctionOutputParameters} consists of many \textit{FunctionParameter}s. 
To enable BPMN tasks to communicate with the smart contract reference (i.e. retrieving data from and writing data to a smart contract), each \textit{SmartContractInterface} can have many \textit{ConnectionOutgoingContractInvocation}s. As aforementioned, interaction with a smart contract is performed via invocations of smart contract functions. Therefore, \textit{ConnectionOutgoingContractInvocation} specifies the signature of the smart contract function to be invoked by the BPMN task via the attribute \textit{fnName}. Each \textit{ConnectionOutgoingContractInvocation} has one of each \textit{FunctionInputBindings} and \textit{FunctionOutputBindings} element. Each of \textit{FunctionInputBindings} and \textit{FunctionOutputBindings} contains multiple \textit{ParameterBinding}s. Using the attribute \textit{values} in \textit{inputParameters} and \textit{outputParameters}, the model can specify bindings from the business process internal variables, or the BPMN task's own input parameters (if it is a user task), to the input parameters and return values of the smart contract function, respectively. 

Taking the grain title record creation as an example, a user can model the grain supply chain business process using BMPN modeller. The \textit{SmartContractInterface} icon can be used to represent the grain title registry smart contract, which is generated based on the grain title registry model built earlier. In the BPMN modeller, the user first needs to click on the grain title registry icon and input the deployed registry contract address for gain title registry. Then the user needs to click on the arrow for invocation of registry action from the BPMN task \textit{Create Grain Title} and bind the input \textit{weight} and \textit{quality} for grain title registry to the respective BPMN process variables \textit{consignmentWeight} and \textit{grainQuality}. 

\subsection{Smart Contract Generation}

Smart contract (SC) generation consists of business process SC translation and registry SC generation. We previously presented our basic business process SC translation algorithm in \cite{Weber:BPM2016,csereport} and basic registry SC generation in \cite{Tran2017}.
In this paper, we extend the SC generation with ERC-20/ERC-721 standard compliance support, and integration between business process execution and registry.



The registry SC generation takes basic information and registry types as fields (e.g., \textit{weight} and \textit{quality}) and operations (e.g., \textit{record\_create()} and \textit{record\_ownership\_transfer()}) as methods in the registry smart contract (e.g., \textit{GrainTitleRegistry} smart contract). To improve the interoperability between Ethereum blockchain applications that work with fungible assets and non-fungible assets, we extend the registry SC generation to produce ERC-20/ERC-721 compliant asset registry smart contracts which follows the rules and methods in the ERC-20\footnote{\url{https://eips.ethereum.org/EIPS/eip-20}} and ERC-721\footnote{\url{https://eips.ethereum.org/EIPS/eip-721}} interface. In other words, the functions in the smart contracts generated by fungible/non-fungible asset registry SC generation contain all the required ERC-20/ERC-721 functions respectively.  

On the business process side, the business process SC translation component takes an existing BPMN model as an input and outputs smart contracts. The output of the BPMN translator includes registry smart contract interfaces (e.g., \textit{GraintTitleRegistry}) and a factory contract which contains information required for instantiating the business process. The factory contract includes the instantiation methods and two types of artifacts, which are an interface specification per role in a business process and a process instance contract. Once deployed, the smart contract functions in the process instance contract can be executed as specified by the BPMN model; the smart contract enforces the process flow, and only conforming instances are possible. For example, we cannot execute \textit{Create Grain Title} before \textit{Grain Weight Evaluated} or \textit{Grain Quality Evaluated} in grain title creation process.  The interface specifications are distributed to the respective triggers while the process instance contracts are deployed to the blockchain when the process instance
is created. The process instance contract contains the implementation of the business logic (e.g., \textit{Create\_grain\_title()}), reflecting the content of the original process specification. 

To support the interaction between business process execution smart contracts and registry smart contracts, SC generation is extended as illustrated in \autoref{algorithm}. For each \textit{SmartContractInterface} element in the business process model, the algorithm produces a smart contract interface in Solidity (e.g., \textit{GrainTitleRegistry} SC interface) with all public functions specified in the element. If the contract address is set in the \textit{SmartContractInterface}, the algorithm generates a variable containing that hard-codes the address for this smart contract interface. Otherwise, the algorithm outputs a process smart contract constructor parameter for setting the referenced smart contract address to a variable. For each \textit{ConnectionOutgoingContractInvocation} element, the algorithm obtains the source task and target smart contract interface from the element and generates Solidity code for invoking the given target smart contract interface function, while mapping inputs and return parameters to process variables as specified. At last, the algorithm injects the generated invocation code to the function body of the respective source task. Listing~\ref{SCCode} presents an example of of generated smart contract \textit{ProcessFactory.sol} for the grain title creation process.

\begin{algorithm}[t] \footnotesize
	\caption{Integration between business processes and registries}
	\label{algorithm}
	\For{each SmartContractInterface element}{
		generate a Solidity smart contract interface with specified public functions\\
		\uIf{contractAddress is set}{
			generate a variable containing hard coded address
		}
		\Else{
			generate a process contract constructor parameter for setting referenced smart contract address to a variable
		} 
	}
	\For{each ConnectionOutgoingContractInvocation element}{
		obtain source task, target smart contract interface\\
		generate Solidity code for invoking given target smart contract interface function while mapping inputs and return parameters to process variables as specified\\
		inject generated invocation code to function body of source task
	}
	
\end{algorithm}

\lstset{  
  frame=single,
  framesep=\fboxsep,
  framerule=\fboxrule,
  xleftmargin=\dimexpr\fboxsep+\fboxrule,
  xrightmargin=\dimexpr\fboxsep+\fboxrule,
  language=Java,
  basicstyle=\footnotesize\ttfamily,
  commentstyle=\color{cyan},
  tabsize=2,
  keywordstyle=,
  breaklines=true,  
  captionpos=b,
  escapeinside=``
}
\begin{lstlisting}[float,caption=Code fragment of a generated smart contract example,label=SCCode]
// -------- EXTERNAL SMART CONTRACT INTERFACES
contract LorikeetCoin {
    function name() external returns (string memory );
    function totalSupply() external returns (uint256 );
    ......
}
contract GrainTitleRegistry {
    function record_get_owner(address record_id) external returns (address record_owner);
    function record_get_attrs(address record_id) external returns (uint256 weight, uint256 quality);
    ......
}
// ----------------------------
contract ProcessFactory {
    function createInstance(address[] memory _participants) public returns(address) {
        ProcessMonitor instance = new ProcessMonitor( /*_participants*/ );
        createdInstances.push(address(instance));
        emit instanceCreated(address(instance));
        return address(instance);
    }
    ......
}
contract ProcessMonitor {
    // ---------- PROCESS VARIABLES
    uint _truckWeightAfter;
    uint _grainWeight;
    ......
    // ----------------------------

    // -------- EXTERNAL SMART CONTRACT ADDRESSES
    address addressOfLorikeetCoin = 0xD3E4EBe81b55EA73b559da31ADf2CAc3b254ea11;
    address addressOfGrainTitleRegistry = 0xA9998dBe75D795556eA821E37cD2DE1F373BFd91;
    // ------------------------------------

    constructor() public {
        _truckWeightAfter = 0;
        _grainWeight = 0;
        ......
    }

    function Create_grain_title(uint preconditionsp) internal returns (uint) {
        if ( (preconditionsp & 0x44 == 0x44) ) {
            GrainTitleRegistry instanceOfGrainTitleRegistry = GrainTitleRegistry(addressOfGrainTitleRegistry);
            instanceOfGrainTitleRegistry.record_create(address(this), _grainWeight, _grainQuality);
            return preconditionsp & uint(~0x44)  | 0x10;
        } else
            return preconditionsp;
    }
    ......
}
\end{lstlisting}



\subsection{Blockchain Interaction}
The approach provides blockchain interaction methods for connecting with a blockchain node, and handling the compilation, deployment as well as communication with smart contracts. Thus, users can monitor the execution of smart contracts and interact with smart contracts directly. Specifically, the communication module comprises sending blockchain tractions, querying smart contract states, and listening to transaction progress and smart contract events. Sending blockchain transactions corresponds to write operations, while querying smart contract states applies to read operations. The communication module obtains status of transactions and receives smart contract events via the listening to transaction progress and smart contract events. 
	
Once the smart contracts are successfully deployed, the users can use the smart contracts to execute the business process instances and create records in fungible/non-fungible asset registries. The approach can furthermore generate user interface (UI) elements to interact with deployed smart contracts. Through these UI elements, users can execute smart contract functions as well as monitor smart contract events. Function invocation user interface forms are automatically populated from smart contract interfaces. Users can retrieve previously emitted events or continuously listen to new events. Process flow conformance is enforced by the approach: if an invocation to execute a task is in conformance with the current process execution state, it takes place; otherwise the invocation is unsuccessful. If a process task was successfully executed, a new event is emitted and the state is updated; if unsuccessful, the information of this invocation is stored but the process state does not change.

\section{MDE Tool: \toolname}
\label{sec:tool}
\begin{figure*}[t]
	\begin{center}
		\includegraphics[width = 0.65\textwidth]{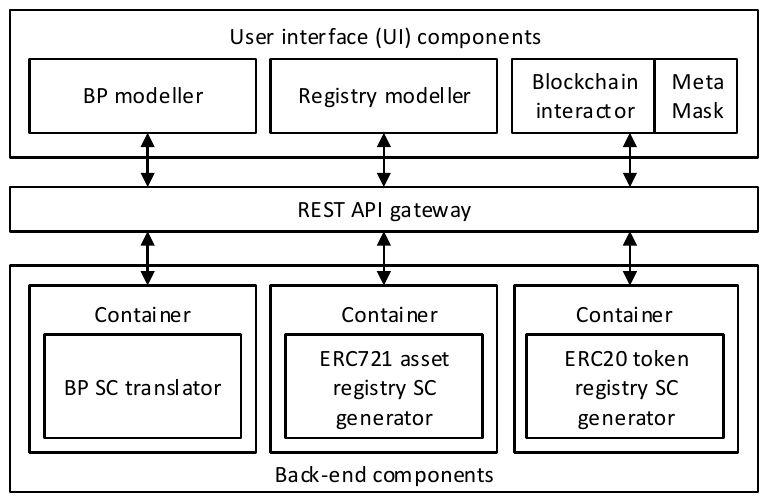}
		\caption{Architecture of Lorikeet.}
		\label{lorikeet}
	\end{center}
\end{figure*}

We design and develop a model-driven engineering tool for business processes and asset management on blockchain, named Lorikeet\footnote{Lorikeet tool demo video: \url{https://drive.google.com/file/d/1rpy-oHbDVkXa6u4Fn73wSX8rINn1sv3U/view} (accessed 26 July 2020)} . As illustrated in \autoref{lorikeet}, the tool consists of the user interface (UI) components and back-end components that are designed adhering to a microservice architecture.

The UI components are presented as web applications for users to build business process and registry models, and interact with the smart contracts. The business process (BP) modeller is extended from the bpmn-js modelling library\footnote{\url{https://github.com/bpmn-io/bpmn-js}}. The user can model fungible/non-fungible asset registry references and action invocation easily by dragging and dropping in the modeller UI and providing the relevant registry information (e.g.\ registry smart contract address, available registry actions, parameter binding information, etc.). The fungible/non-fungible asset registry modeller provides a form for users to fill in the information required by the registry model.
The users can compile, deploy, and interact with the smart contracts in the blockchain interactor. The supported interactions include deployment and execution of smart contract functions as well as monitoring of smart contract events. The smart contract function invocation UI forms are automatically populated from the respective smart contract interface, while the smart contract event monitoring UI can display all previously emitted events or list any events by continuously listening to the blockchain. 
The blockchain interactor component is written in TypeScript with Node.js version 10, implementing the REST API using express.js server. Users can securely create and manage their identities via MetaMask\footnote{\url{https://metamask.io/}} which connects to the blockchain interactor.

Back-end components, including business process (BP) smart contract (SC) translator, ERC721 asset registry SC generator, and ECR20 token registry SC generator, are built and deployed independently as Docker containers. BP SC translator automatically generates smart contracts from BP models, while ERC721 asset/ERC20 token registry generator derives smart contracts based on the registry models. 
For Ethereum, the smart contracts are written in Solidity, compiled with Solidity compiler version 0.4.24. We used Truffle framework v12 to compile and test smart contracts. 

The BP/registry modeller and blockchain interactor communicate with the back-end microservices via an API gateway. The API gateway forwards API calls from the UI, such as translating BPMN model to smart contract code, to the corresponding microservice. In addition, the blockchain interactor sends information about the emitted events in real-time using socket.io.

\begin{figure*}[t]
	\begin{center}
		\includegraphics[width = \textwidth]{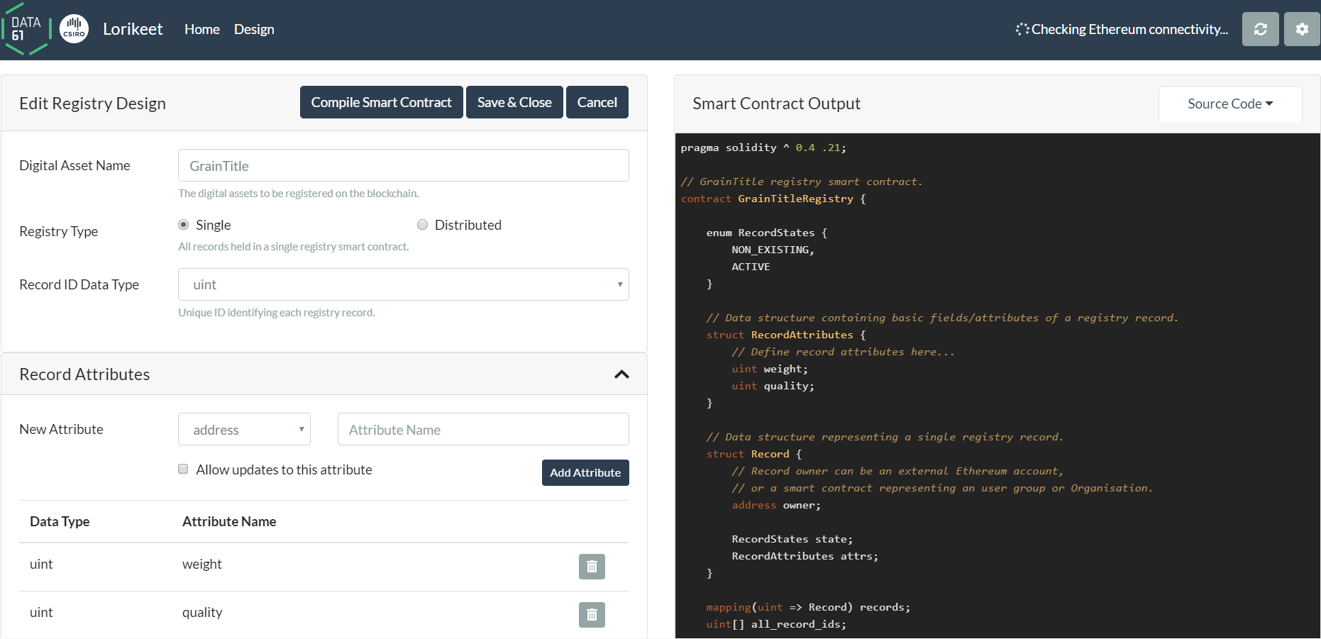}
		\caption{Graphical interface of \toolname.}
		\label{fig:tool-ui}
	\end{center}
\end{figure*}

\autoref{fig:tool-ui} shows the business process modeller UI of our tool. Once the user makes changes to the BP model on the left-hand side, it is translated to the corresponding smart contract code which is displayed on the right-hand side. The UIs for ERC721 asset/ERC20 token registry modelling are similar to the BP modeller UI: in both cases there is a form on the left-hand side collecting the customised registry information.

\section{Evaluation}
\label{sec:eval}
To evaluate the feasibility, functional correctness, and cost of our model-driven engineering approach, we use Lorikeet to model four use cases and generate smart contract code based on the models. The selected use cases include initial coin offering (ICO) process, quality tracing process, task outsourcing process, and grain title creation process. They cover fungible/non-fungible asset registration, escrow for conditional payment, and asset swap respectively. We encountered these use cases in projects or discussions with industry or government, and modelled them based on publicly available information. According to the classification of experiments by Zelkowitz and Wallace \cite{Zelkowitz1998}, this evaluation falls into the category of case studies since we do not have control over the experimental conditions: the processes are taken from external units (industry or government). The selected use cases are modelled as cross-organisational processes in a single pool, following the process model design philosophy for blockchain-based processes outlined in \cite{Lopez-Pintado:BPM2017}.

\begin{figure*}[t]
	\begin{center}
		\includegraphics[width = 0.9\textwidth]{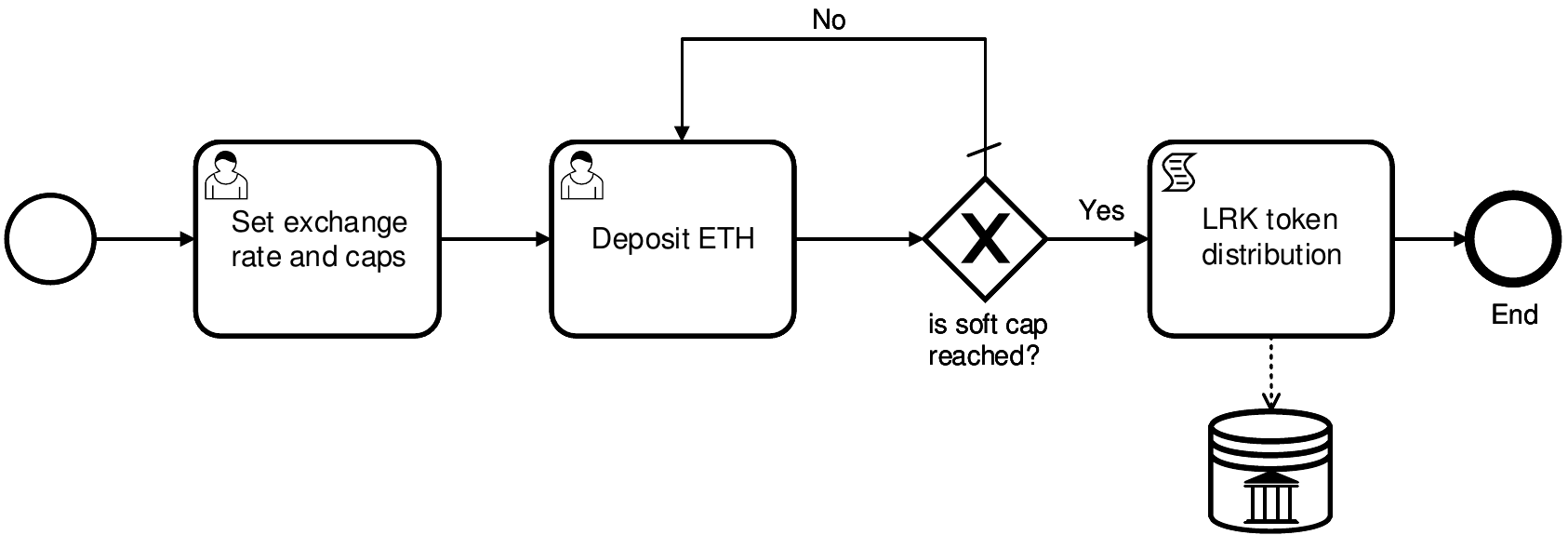}
		\caption{ICO process model built using Lorikeet.}
		\label{case-ico}
	\end{center}
\end{figure*}

\subsection{Feasibility}
In this section, we evaluate the feasibility of our MDE approach using four uses, including ICO process (Section 5.1.1), quality tracing process (Section 5.1.2), task outsourcing process (Section 5.1.3), and grain title creation process (Section 5.1.4).

\subsubsection{ICO Process}
\label{sec:ico}
An Initial Coin Offerings (ICO) is a way to raise funds for new projects using cryptocurrency: essentially, investors can buy tokens for a new startup / initiative / project and pay for those using an established cryptocurency like Ether or Bitcoin. 
ICOs can be considered as fungible asset registration use cases, where the new token is the fungible asset.
\autoref{case-ico} shows an ICO business process model built using Lorikeet. The process starts with setting up the token exchange rate with Ether (i.e. ETH) and caps (including one soft cap and one hard cap). Once the token exchange rate and caps are determined, investors deposit ETH to the business process smart contract.
If the soft cap is reached, the bought amount of tokens are transferred to the investors' accounts, which are recorded in the token registry.
The registry smart contract address and actions to be invoked for the registry are specified as attributes of \textit{SmartContractInterface}. On the registry side, we built a fungible asset registry model for the token registry by filling the form provided by Lorikeet. The token name is \textit{Lorikeet Coin}, while the symbol is \textit{LRK}. The decimal value we put is \textit{2}. 

Lorikeet creates a smart contract for instantiating the ICO process taking the business process model as input, while generating a token registry smart contract using the specified token information in the form as fields and the ERC-20 standard methods as methods. The token registry smart contract interface defines the methods interacting with the corresponding token registry smart contract while the process monitor smart contract implements the business process instance. 

The results show that Lorikeet can automatically generate an ICO business process smart contract and a ERC-20 standard compliant ICO token registry smart contract using the ICO BPMN model and token registry data schema respectively. The output business process smart contract checks whether business process instances run compliant with the BPMN model (e.g. only when soft cap is reached, LRK tokens are distributed), while the token registry smart contract maintains the tokens in each investor's account. Instead of writing those two complex smart contracts, developers only need to focus on designing and refining high-level models, which are easier to explain to ICO stakeholders and to check for correctness. For example, the developers only need to specify the initially distributed accounts by filling the respective field in the fungible asset registry form provided by Lorikeet for LRK token distribution. Also, as the token registry smart contract is ERC-20 compliant, it can be easily integrated with other Ethereum applications following the ERC-20 standard. The results show that Lorikeet can successfully support MDE of business processes and registries concerning fungible assets.

\subsubsection{Quality Tracing Process}
\label{sec:quality}
\begin{figure*}[t]
	\begin{center}
		\includegraphics[width = 0.9\textwidth]{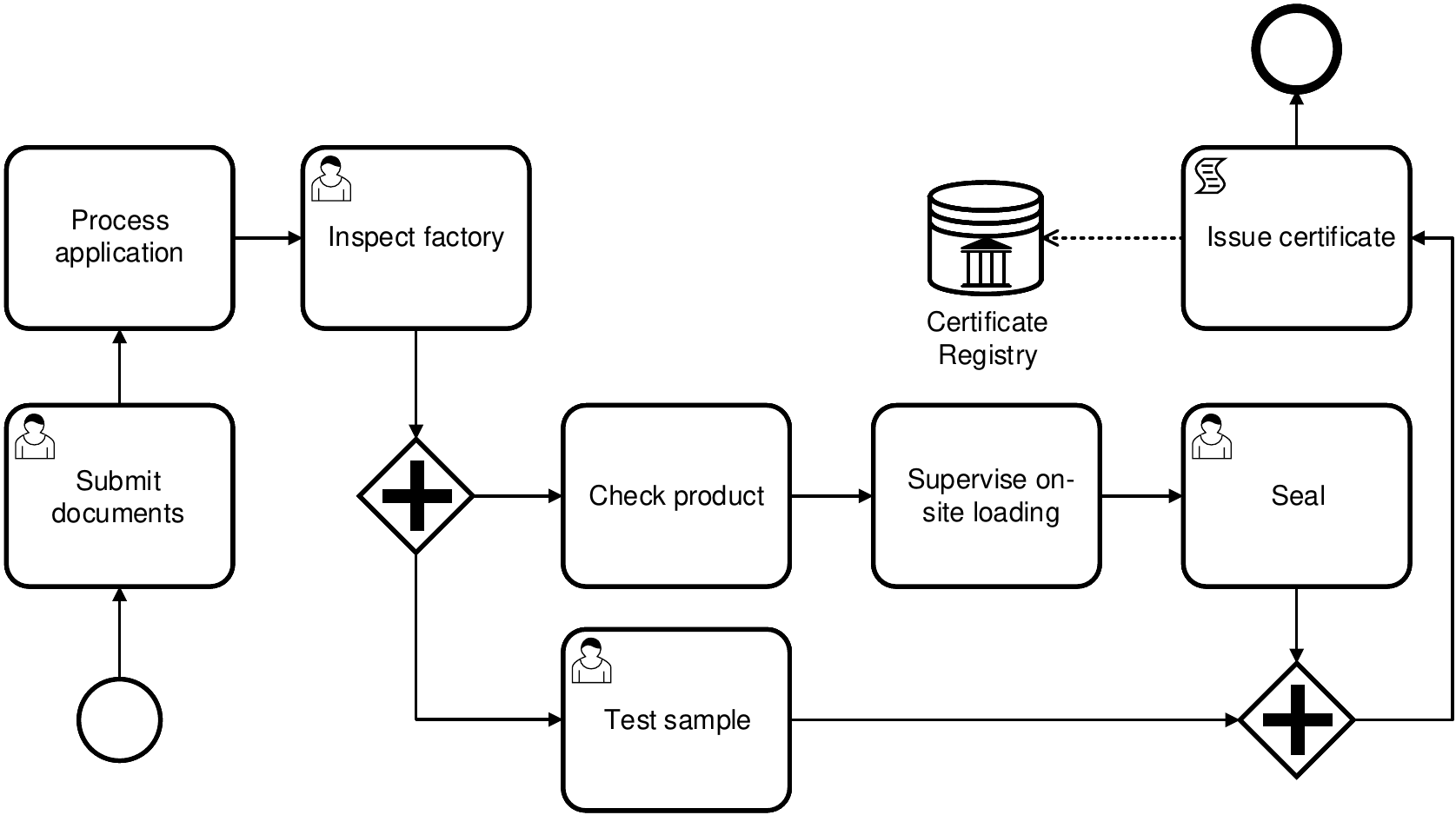}
		\caption{Quality tracing process model built using Lorikeet.}
		\label{case-quality}
	\end{center}
\end{figure*}

\autoref{case-quality} illustrates the quality tracing process models for import commodities in China \cite{qualitytracing} built using Lorikeet, which can be viewed as a non-fungible asset registration use case. The quality inspection agency provides quality tracing services and issues a traceability certificate of commodity if all requirements are fulfilled. The process starts when a product supplier lodges a quality tracing application for each batch of products to the quality inspection company. The administrator processes the paper work (e.g.\ invoices) and payment. Then the agency assigns a factory examiner to check the factory address, production capability, quality control process, etc. After inspecting the factory, a freight yard examiner is sent to check the products on freight yard and inspect on-site loading. The examiner attaches lead seals to the product containers if the on-site loading processes meet requirements. In the meantime, a product sample is sent to a lab for sample testing. Once the application passes the inspections and testing, the agency issues the supplier a traceability certificate of commodity. All the relevant traceability information and certificates are stored in the traceability registry.

The quality tracing process and traceability registry are modelled in a similar way as ICO use case discussed in \autoref{sec:ico}. 
The registry smart contract address and actions to be invoked for the registry are specified as attributes \textit{SmartContractInterface}. On the registry side, we built a non-fungible asset registry model for the certificate registry by filling the Lorikeet registry template. The asset name is \textit{CertificateOfOrigin} and the registry type is \textit{single}. The record ID represents certificateID. The attributes in the registry model include \textit{factoryReport}, \textit{testReport}, and \textit{freightyardReport}, which store the hash values of respective documents.

Lorikeet outputs a smart contract for creating a quality tracing process instance based on the built business process model and generates a ERC-721 compliant smart contract for the traceability certificate registry based on the built registry model. 

The generated quality tracing process smart contract can check whether the process instances are executed consistently with the BPMN models (e.g. factory inspection, sample tests, and on-site loading supervision must be done before issuing a certificate), while the certificate registry smart contract maintains the hash values of certificates. Developers only need to specify a quality tracing process model and certificate registry data schema to achieve a smart contract implementation. Also, other blockchain applications can easily interact with the generated certificate registry smart contract since it follows the ERC-721 standard. These results show that Lorikeet can provide efficient MDE support of non-fungible asset business processes and registries.


\begin{figure*}[t]
	\begin{center}
		\includegraphics[width = 0.95\textwidth]{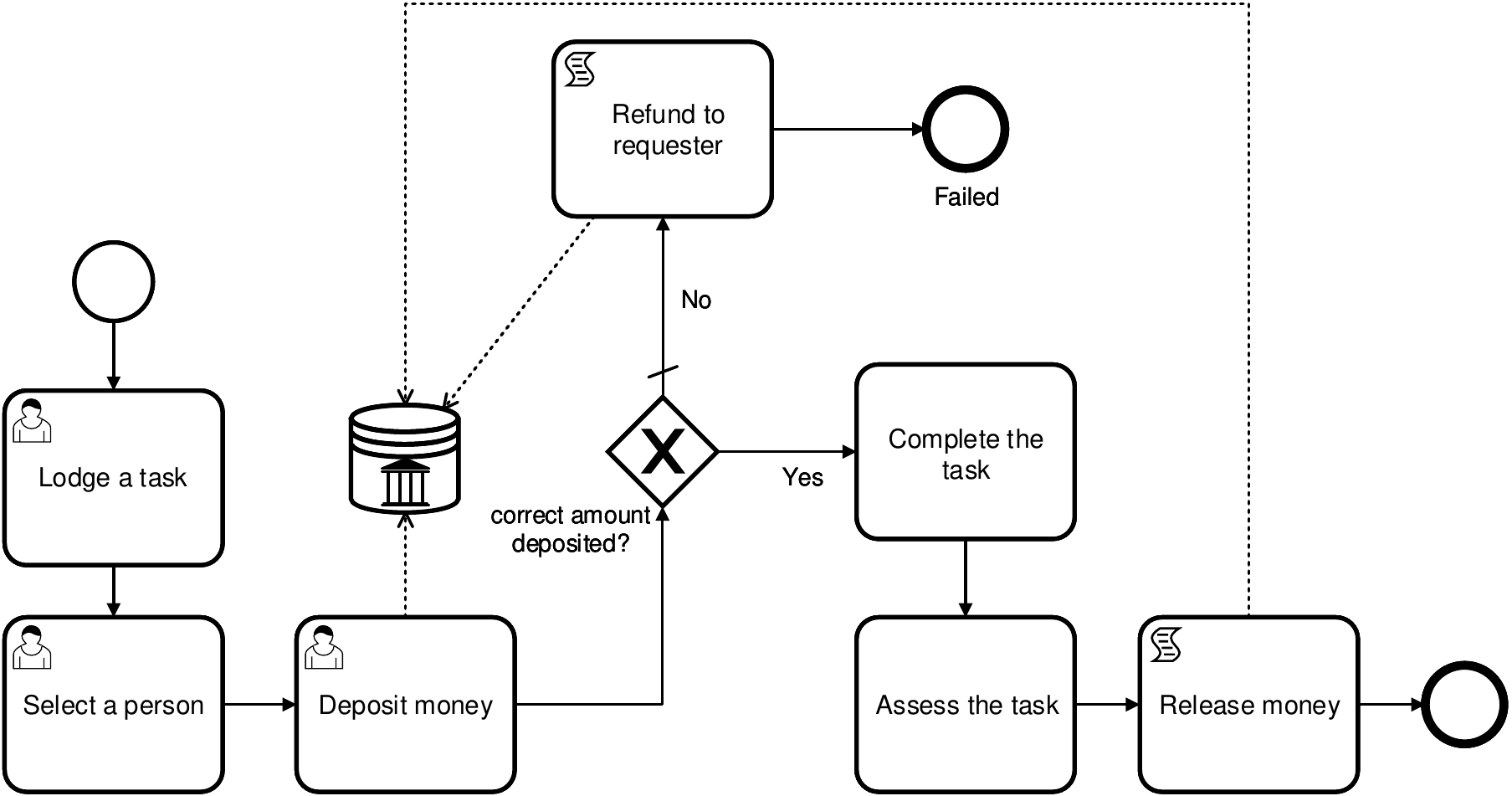}
		\caption{Task outsourcing process model built using Lorikeet.}
		\label{case-task}
	\end{center}
\end{figure*}

\subsubsection{Task Outsourcing Process}
\label{sec:task}

\autoref{case-task} illustrates a task outsourcing business process model, which is an escrow use case for conditional payment. The process starts when a task requester lodges a task. The task requester selects a person from a list of matching workers and deposits the negotiated amount of money to the token registry. If the amount of deposited money is not correct or the recipient is wrong, the money is refunded back to the task requester. Otherwise, the money is released to the worker if the task is assessed as complete. The task outsourcing process model specifies the Lorikeet token registry using \textit{SmartContractInterface}. On the registry side, we use the same fungible asset registry model for the Lorikeet token registry. 

Lorikeet creates a smart contract for instantiating the task outsourcing process, taking the business process model as input while generating a token registry smart contract using the specified token information as fields and the ERC-20 standard methods as methods. 

This use case shows that conditional payment (i.e., escrow) can be implemented using Lorikeet. Developers only need to model the payment process and token registry without writing Solidity code for  complex escrow logic.


\begin{figure}[!ht]
	\begin{center}
		\includegraphics[width = \textwidth]{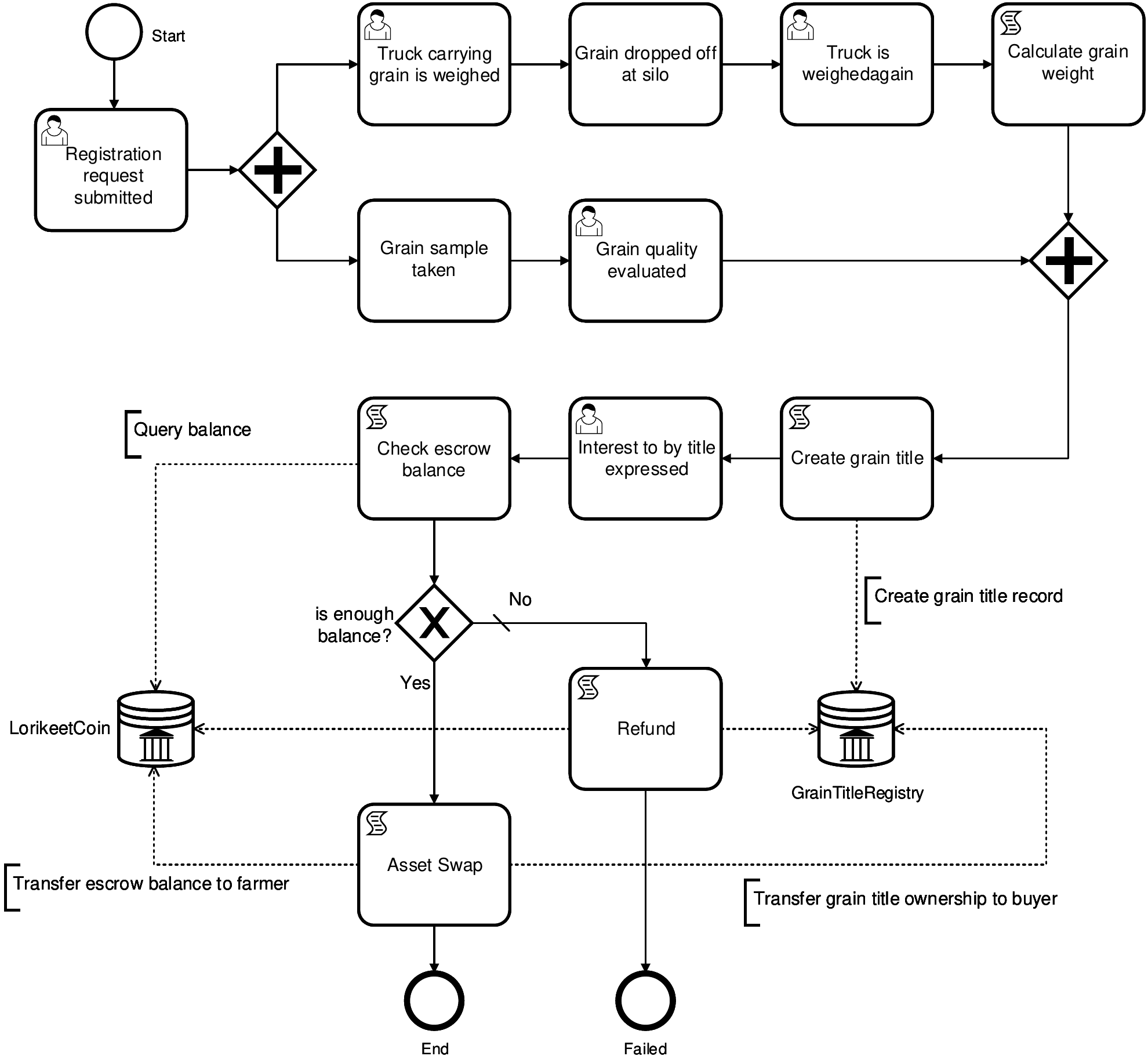}
		\caption{Grain title creation process model built using Lorikeet.}
		\label{case-grain}
	\end{center}
\end{figure}

\subsubsection{Grain Title Creation}
\label{sec:grain}
\autoref{case-grain} shows a simplified grain title creation process modelled using Lorikeet, which is focused on grain ownership transfer \cite{aureport} and can be considered an asset swap use case. There are two asset registries interacting with this selected grain title creation process: grain title registry (non-fungible) and Lorikeet token registry (fungible). 

The grain title creation process starts when a truck arrives and is weighed for the first time. The 
grain is dropped into a silo and the truck is weighed again to determine the net weight of grain that has been delivered (weight of truck before delivery minus weight of truck after delivery). Also, when the truck arrives, before dropping grain into the silo, a grain sample is taken and quality assessment is conducted. The process then creates the grain title and assigns it to the farmer, but puts it into escrow of the process. When the buyer pays the correct amount of money for the grain into the process escrow, process transfers the grain title to the buyer and the money to the farmer. If the amount is incorrect the money and the title are refunded to the buyer and the farmer, respectively.

The registry smart contract address and actions to be invoked for the grain title registry are specified as attributes of \textit{SmartContractInterface}. 
On the registry side, we use the same fungible asset registry model for the Lorikeet token registry. 


Lorikeet outputs a smart contract for the grain title creation process taking the business process model as input. For registry smart contracts, Lorikeet uses the specified token and grain title information to generate an ERC-20 compliant smart contract and an ERC-721 compliant smart contract, which can be easily integrated and communicated with other ERC smart contracts. The process of grain title creation involves various activities and interactions with both fungible asset registry and non-fungible asset registry. 

When using Lorikeet, developers only need to model the process and the two asset registries without writing Solidity code for complex business logic, such as creating grain title, checking escrow balance, and asset swap. Also, the business process implemented as a smart contract enforces conformance of any execution with the grain title process, e.g., a grain title can only be created after grain weight is calculated and quality is evaluated.

\subsection{Functional Correctness and Cost Analysis}

\begin{table*}[t]
	\footnotesize
	\centering
	\caption{Grain title creation process conformance checking results}
	\label{tab:conformance}
	\smallskip
	\begin{tabular}{p{0.12\columnwidth}p{0.12\columnwidth}p{0.20\columnwidth}p{0.15\columnwidth}p{0.15\columnwidth}}
		\toprule
		
		\multicolumn{1}{l}{Tasks} &
		\multicolumn{1}{l}{Gateways} &
		\multicolumn{1}{l}{Trace type} &
		\multicolumn{1}{l}{Traces} &
		\multicolumn{1}{l}{Correctness} \\
		\midrule
		
		\multirow{2}{0.12\columnwidth}{12} & \multirow{2}{0.12\columnwidth}{3} & Conforming & 77 & 100\% \\
		\cmidrule(l){3-5}
		
		&  & Not conforming & 425 & 100\% \\
		
		\bottomrule
	\end{tabular}
\end{table*}

We tested the functional correctness of Lorikeet by checking conformance of the generated business process, including interactions between the registry smart contracts and the business process smart contracts, with a test suite in the Given-When-Then structure~\cite{gottesdiener2012discover}. We ran the experiment on \emph{Ganache}, i.e. the blockchain client simulates an Ethereum blockchain, and compiled all the Solidity smart contracts using \emph{solc} v.0.5.8 with optimization enabled. 

In our earlier work \cite{Weber:BPM2016}, we derived the set of permissible execution traces for each process model and randomly modified these traces to obtain a larger set of not conforming traces. Then we tested the ability of the smart contracts to discriminate between correct and incorrect traces, which it did perfectly. The control flow logic has not changed, and thus there was no need to rerun the experiments for this paper. 

In order to provide a concrete example, we used the grain title creation process as shown in \autoref{case-grain}. We started with deriving two permissible execution traces for the grain title creation process model, so-called \emph{conforming traces} that adhere to the process model. One follows the successful \textit{Asset Swap} path, while the other follows the failed \textit{Refund} path. For each of these two traces, we generated 250 traces with randomized noise injected to obtain a larger set of traces (including both conforming and non-conforming traces) with the following manipulation operators: (i) add a new log line, (ii) remove a log line, or (iii) switch the order of two log lines, such that the modified trace was different from the two initial conforming traces. In total, there are 502 traces. 

We investigated if our implementation accurately identifies the non-conforming traces that have been generated for the grain title creation process model. The script tasks are executed with the preceding task. In the current version of Lorikeet, function invocation UI forms are automatically populated from the smart contract interface. The functions can be called by sending the transaction with the click of a button on the UI. The results are shown in \autoref{tab:conformance}. All log traces were correctly classified. There are no assertion failures, which was our expectation. Any other outcome (i.e. assertion failure) would have pointed at severe issues with our approach or implementation. Thus, we claim if a trace is conforming, its execution will execute the correct logic that we expect. Our full test suite is available online\footnote{\url{https://drive.google.com/drive/folders/1gaWhCY2YK8n4MXUvoA69jzbNboJ8iTQZ?usp=sharing}}.

We also investigated the cost of involving the blockchain in the process execution, since 
gas cost on Ethereum reflects computational effort and determines throughput on a given network. 
Each bytecode instruction in a smart contract consumes gas when it is executed in response to a transaction invoking the smart contract.
The ``size'' of a block in Ethereum is specified as a block gas limit, i.e., the sum of gas consumed by all transactions in a block may not exceed this limit. 
As such, cost in gas affects all blockchain networks, regardless of whether they are public, private, or consortium blockchain networks.
\begin{table*}[t]
	\footnotesize
	\centering
	\caption{Gas used by each task in the grain title creation process}
	\label{tab:cost}
	\smallskip
	\begin{tabular}{p{0.45\columnwidth} r r r}
		\toprule
		
		\multicolumn{1}{l}{Task} &
		\multicolumn{1}{r}{Avg} &
		\multicolumn{1}{r}{Min} &
		\multicolumn{1}{r}{Max} \\
		\midrule
		
		GrainTitleRegistry\_deployment & 803278 & 803278 & 803278 \\
		\cmidrule(l){1-4}
		
		LorikeetCoin\_deployment & 1486678 & 1486678 & 1486678 \\
		\cmidrule(l){1-4}
		
		ProcessMonitor\_deployment & 1359815 & 1359815 & 1359815 \\
		\cmidrule(l){1-4}
		
		Registration\_request\_submitted & 49491 & 49491 & 49491 \\
		\cmidrule(l){1-4}
		
		Truck\_carrying\_grain\_is\_weighed & 49319 & 49319 & 49319 \\
		\cmidrule(l){1-4}
		
		Grain\_sample\_taken & 28983 & 28983 & 28983 \\
		\cmidrule(l){1-4}
		
		Grain\_dropped\_at\_silo & 28938 & 28938 & 28938 \\
		\cmidrule(l){1-4}
		
		Grain\_quality\_evaluated & 93055 & 49296 & 146411 \\
		\cmidrule(l){1-4}
		
		Truck\_is\_weighed\_again & 102321 & 69892 & 182007 \\
		\cmidrule(l){1-4}
		
		Interest\_to\_buy\_title\_expressed & 109523 & 104903 & 119986 \\
		\cmidrule(l){1-4}
		
		Failed & 14338 & 14338 & 14338 \\
		\cmidrule(l){1-4}
		
		End & 14349 & 14349 & 14349 \\
		
		\bottomrule
	\end{tabular}
\end{table*}

\autoref{tab:cost} shows the average, maximum, and minimum value of gas used by deployment and each activity. The first three rows show the deployment cost of the three generated smart contracts. Note that the variability in gas cost (difference between minimum and maximum) used by Grain\_quality\_evaluated and Truck\_is\_weighed\_again is large. The reason is that these two tasks are (eventually) followed by a parallel join gateway and thereafter a script task; only once the second of the two tasks completes, the gateway and subsequent script task are executed. Therefore, the cost depends on the execution order, resulting in the high variability.
Also noteworthy is that Interest\_to\_buy\_title is consistently expensive. It too is followed by script tasks and a gateway, but regardless of the gateway decision there are always two scripts executed. And all of the involved script tasks interact with the asset registries.
Finally, both Failed and End are end events of the BPMN process, and the generated Solidity code for both involves re-setting a smart contract storage variable.
This triggers a refund of 15,000 gas, which offsets the total gas consumption of the rest of the transaction, and hence results in a low total cost.

The average gas cost per transaction an instance of the process model is 59,497 gas (taking into account that only one of the outgoing branches of the XOR split is executed).
For comparison, we used Google BigQuery's Ethereum dataset\footnote{\url{https://cloud.google.com/blog/products/data-analytics/ethereum-bigquery-public-dataset-smart-contract-analytics}} to determine the average cost of a transaction that invokes a smart contract, with the following query:
\lstset{  
  language=SQL,
  showspaces=false,
		       basicstyle=\footnotesize\ttfamily, 
  commentstyle=\color{gray}
}
\begin{lstlisting}[]
SELECT avg(receipt_gas_used) 
  FROM 'bigquery-public-data.ethereum_blockchain.transactions'
  WHERE DATE(block_timestamp) <= "2019-12-11" 
    AND to_address is not null 
    AND to_address in 
      (SELECT address 
        FROM 'bigquery-public-data.ethereum_blockchain.contracts') 
    AND input != "0x" 
\end{lstlisting}
From this query, we obtained aggregate costs over 292,826,434 contract invocation transactions on the public Ethereum blockchain up until 11 December 2019: on average 86,330 gas; median of approx. 45,647.\footnote{The median is approximate, since the large total number requires using approximate SQL functions like \texttt{APPROX\_QUANTILES}.}
By comparison, transactions invoking functions of the smart contracts generated with Lorikeet use on average only 59,497 gas, or 68.9\% of the global average; the median is 49,405 gas, which is about 8\% higher than the global median.

\section{Conclusion and Future Work}
\label{sec:concl}
Model-driven engineering (MDE) is of particular importance for blockchain-based applications since MDE tools can help developers focus on high-level modelling and generate well-tested smart contract code implementing best practices. A typical class of decentralised applications uses blockchain to manage business processes that interact with asset registries, including processes for fungible/non-fungible asset registration, escrow for conditional payment, and asset swap. This paper tackled the integration of business processes and asset management, which requires modelling support for various types of integrations, and smart contract generation based on the build models.  

We proposed methods to specify models for business processes and asset registries, to interconnect them, and to generate smart contracts using the specified models. To support the proposed MDE approach, we designed and implemented a tool named Lorikeet. The proposed approach was evaluated in terms of feasibility and functional correctness using four industrial use cases. Code from the evaluation has been made available for reproducibility. The results show that developers can use our MDE approach as implemented in the Lorikeet tool to generate functionally correct smart contracts based on the business process and asset registry models. 

A comparison of the gas consumption of a transaction to the generated smart contracts vs.\ over more than 292 million contract invocation transactions on public Ethereum shows that the former consumes less gas on  average, but at a slightly higher median.
While this data is not suitable for drawing definitive conclusions from a comparison of absolute numbers -- after all, we do not know what functions are implemented by other smart contracts -- they give strong indication that Lorikeet smart contracts are not overly inefficient. Note that lower gas consumption corresponds to higher throughput and, on public blockchains, lower monetary cost -- therefore gas consumption is relevant for private and consortium blockchains as well.

Although we focus on the domain of business processes and asset management in this paper, our approach can be easily applied to a broad range of blockchain applications. In future work, we plan to extend our MDE approach to support more comprehensive access control policies across all types of generated smart contracts.

\printendnotes



\end{document}